\documentclass[fleqn,usenatbib,useAMS]{mnras}

\usepackage[T1]{fontenc}

\DeclareRobustCommand{\VAN}[3]{#2}
\let\VANthebibliography\thebibliography
\def\thebibliography{\DeclareRobustCommand{\VAN}[3]{##3}\VANthebibliography}

\usepackage{graphicx}	
\usepackage{amsmath}	
\usepackage{amssymb}	

\usepackage{newtxtext,newtxmath}

\newcommand{\msunyr}{\textrm{ M}_\odot\textrm{ yr}^{-1}}

\title[Nova irradiation]{Novae heat their food: mass transfer by irradiation}

\author[S. Ginzburg and E. Quataert]{
Sivan Ginzburg$^{1}$\thanks{E-mail: ginzburg@berkeley.edu}\thanks{51 Pegasi b Fellow.}
and Eliot Quataert$^{2}$
\\
$^{1}$Department of Astronomy and Theoretical Astrophysics Center, University of California, Berkeley, CA 94720, USA\\
$^{2}$Department of Astrophysical Sciences, Princeton University, Princeton, NJ 08544, USA
}

\date{Accepted XXX. Received YYY; in original form ZZZ}

\pubyear{2021}

\begin{document}
\label{firstpage}
\pagerange{\pageref{firstpage}--\pageref{lastpage}}
\maketitle

\begin{abstract}
A nova eruption irradiates and heats the donor star in a cataclysmic variable to high temperatures $T_{\rm irr}$, causing its outer layers to expand and overflow the Roche lobe. We calculate the donor's heating and expansion both analytically and numerically, under the assumption of spherical symmetry, and find that irradiation drives enhanced mass transfer from the donor at a rate $\dot{m}\propto T_{\rm irr}^{5/3}$, which reaches $\dot{m}\sim 10^{-6}\msunyr$ at the peak of the eruption --- about a thousand times faster than during quiescence. As the nova subsides and the white dwarf cools down, $\dot{m}$ drops to lower values. We find that under certain circumstances, the decline halts and the mass transfer persists at a self-sustaining rate of $\dot{m}\sim 10^{-7}\msunyr$ for up to $\sim 10^3$ yr after the eruption. At this rate, irradiation by the white dwarf's accretion luminosity is sufficient to drive the mass transfer on its own. The self-sustaining rate is close to the white dwarf's stable burning limit, such that this bootstrapping mechanism can simultaneously explain two classes of puzzling binary systems: recurrent novae with orbital periods $\approx 2$ h (T Pyxidis and IM  Normae) and long-lived supersoft X-ray sources with periods $\approx 4$ h (RX J0537.7--7034 and 1E 0035.4--7230). Whether or not a system reaches the self-sustaining state is sensitive to the donor's chromosphere structure, as well as to the orbital period change during nova eruptions.  
\end{abstract}




\section{Introduction}

Cataclysmic variables (CVs) are binary systems in which a donor star (we focus here on main sequence donors) fills its Roche lobe and transfers mass to a white dwarf. As hydrogen accumulates on the white dwarf's surface, the density and temperature at the base of the accreted layer rise. At some point, a critical mass is reached, and the accumulated layer explodes in a thermonuclear runaway --- a nova. The binary system survives the explosion, mass accumulates once again on top of the white dwarf, and the nova recurs on a time-scale that depends on the accretion rate and on the white dwarf's mass \citep[e.g.][]{GallagherStarrfield78,Shara89,PrialnikKovetz95,TownsleyBildsten2004,Yaron2005, Wolf2013,Chomiuk2020}. 

During and immediately after a nova eruption, the white dwarf's luminosity is close to the Eddington limit \citep{Prialnik86,Gehrz98}. The companion is irradiated and heated to a surface temperature that is an order of magnitude hotter than its pre-eruption (main sequence) effective temperature. \citet{Kovetz88} studied the response of the companion to such irradiation and calculated the heat penetration into its atmosphere. As the heated layers expand, the donor overflows its Roche lobe more than in quiescence, and the mass transfer rate increases by orders of magnitude. Recently, \citet{Hillman2020} demonstrated that this irradiation-driven enhanced mass transfer may dominate the long term evolution of CVs over multiple nova cycles. 

Here, we revisit the heating of the donor star by the hot white dwarf during and after a nova eruption, under the simplifying assumption of spherical symmetry, using a combination of analytical arguments and experiments with the \textsc{mesa} stellar evolution code \citep{Paxton2011,Paxton2013}. We calculate the enhanced mass transfer rate following a nova and improve upon \citet{Kovetz88} and \citet{Hillman2020} by refining their power-law scaling for how this rate depends on the irradiation temperature and on the donor's mass. More importantly, we revise upwards the normalization of the mass transfer rate.

Finally, we discuss the decline of the mass transfer rate back to quiescence as the white dwarf cools down after a nova. We find that under certain conditions, the decline halts and the mass transfer becomes self-sustaining: the white dwarf's accretion luminosity is enough to drive the accreted mass through the irradiated companion's Roche lobe. This `bootstrapping' state may explain the high quiescent accretion rates ($\sim 10^{-7}\msunyr$) recently observed for the recurrent novae T Pyxidis (T Pyx) and IM Normae (IM Nor) \citep{Patterson2017,Patterson2020,Godon2018}. If the self-sustaining rate is sufficiently high, the white dwarf can burn accreted hydrogen stably \citep{Nomoto2007,ShenBildsten2007,Wolf2013}, potentially explaining long-lived supersoft X-ray sources with short orbital periods like RX J0537.7--7034 and 1E 0035.4--7230 as well \citep{KahabkaVanDenHeuvel1997, King2001}. 

The remainder of this paper is organized as follows. In Section \ref{sec:quiescence} we analyse the Roche-lobe filling during quiescence, and in Section \ref{sec:eruption} we study the response of the companion to a nova outburst. Our main results --- the enhanced mass transfer rates --- are presented in Section \ref{sec:rates}, where we also compare to previous works (Section \ref{sec:previous}) and discuss the self-sustaining bootstrapping state (Section \ref{sec:boots}). We summarize our findings in Section \ref{sec:summary}.   

\section{Quiescence}\label{sec:quiescence}

Between nova eruptions, the binary loses angular momentum by magnetic braking and gravitational waves on a $\sim$ Gyr time-scale, and the companion (donor) star is driven towards stable Roche-lobe overflow \citep[e.g.][]{Rappaport1983,SpruitRitter83,Knigge2011}. The companion, with a mass $m$ and a radius $r$, overflows its Roche lobe $r_{\rm L}$ by $\Delta r\equiv r-r_{\rm L} \ll r$ such that the mass transfer rate $\dot{m}$ through the L1 Lagrange point follows the angular momentum loss rate \citep[up to an order-unity factor; see][]{Rappaport82} $\dot{m}\sim 10^{-10}-10^{-9}\msunyr$.\footnote{Throughout the paper, $\dot{m}$ is to be understood as $|\dot{m}|$, where we omit the absolute value sign for brevity.} Note that the photosphere can be either interior or exterior to the Roche lobe. 

We estimate the overfilling length $\Delta r$ using the standard calculation, in which the overflowing companion's surface is modelled as a sphere, truncated at a depth $\Delta r$ by the Roche potential \citep[e.g.][]{PaczynskiSienkiewicz72,Savonije78,Ritter1988,KolbRitter1990,Marsh2004,LinialSari2017}. The mass flow rate through the L1 nozzle is given by \citep[e.g.][]{Ritter1988} 
\begin{equation}\label{eq:m_dot}
    \dot{m}\approx\frac{2\upi F(q)}{\sqrt{\rm e}}\rho c_{\rm s}r\Delta r,
\end{equation}
where $\rho$ and $c_{\rm s}$ are the density and sound speed at depth $\Delta r$ inside the companion's atmosphere, respectively \citep[the flow reaches sonic velocities at L1 and the density is reduced by a factor of $\sqrt{\rm e}$ when applying Bernoulli's equation along a streamline; see][]{LubowShu75,Ritter1988}. The nozzle's cross section is $\upi[r^2-(r-\Delta r)^2]\approx 2\upi r \Delta r$, up to a correction $F$ that depends logarithmically on the primary to secondary mass ratio $q$ \citep[in our case $F\approx 1.5$; see][]{Ritter1988}.
We normalize $\dot{m}$ using
\begin{equation}\label{eq:m_ph}
    \dot{m}_{\rm ph}\equiv \dot{m}(\Delta r\approx h)\approx\frac{2\upi F}{\sqrt{\rm e}}\left(\frac{kT_{\rm eff}}{\mu}\right)^{3/2}\frac{r^3}{Gm}\rho_{\rm ph},
\end{equation}
which is the mass loss rate of companions that overflow their Roche lobe by a photospheric scale height $h=kr^2T_{\rm eff}/(Gm\mu)$, where $G$, $k$, and $\mu$ denote the gravitational constant, Boltzmann's constant, and the molecular weight, respectively; $T_{\rm eff}$ is the donor's effective temperature. We estimate the photospheric density $\rho_{\rm ph}$ using the condition
\begin{equation}
    \kappa \rho_{\rm ph} h\sim 1,
\end{equation}
where the opacity $\kappa\propto\rho_{\rm ph}^{1/2}T_{\rm eff}^9$ is dominated by ${\rm H}^-$ for typical M dwarf companions with $T_{\rm eff}\sim 3\times 10^3\textrm{ K}$ \citep[e.g.][]{Hansen2004}. We substitute $\rho_{\rm ph}$ in equation \eqref{eq:m_ph} and find \citep[see also][]{Ritter1988}
\begin{equation}
\dot{m}_{\rm ph}\propto r^{5/3}m^{-1/3}T_{\rm eff}^{-31/6}\sim 10^{-8}\msunyr.    
\end{equation}
We rewrite equation \eqref{eq:m_dot} as
\begin{equation}\label{eq:t_loss}
\dot{m}=\dot{m}_{\rm ph}\frac{\rho}{\rho_{\rm ph}}\left(\frac{T}{T_{\rm eff}}\right)^{1/2}\frac{\Delta r}{h},
\end{equation}
where $T$ is the temperature at depth $\Delta r$.

If the donor overflows the Roche lobe by $\Delta r\gg h$, its structure can be approximated by a polytrope (adiabat) with an index $n$, which terminates at a well-defined stellar surface:
\begin{equation}\label{eq:polytrope}
    \frac{\rho}{\rho_{\rm ph}}\sim\left(\frac{T}{T_{\rm eff}}\right)^n\sim\left(\frac{\Delta r}{h}\right)^n.
\end{equation}
If, on the other hand, the donor's photosphere {\it underfills} the Roche lobe by several scale heights (i.e. $\Delta r<0$), then
the profile is isothermal with $T\approx T_{\rm eff}$, and the density $\rho$ falls off exponentially from the photosphere towards L1 \citep{Ritter1988} such that 
\begin{equation}\label{eq:t_loss_general}
    \frac{\dot{m}}{\dot{m}_{\rm ph}}\sim\begin{cases}
    (\Delta r/h)^{n+3/2} & +\Delta r\gg h \\
    \exp{(\Delta r/h)} & -\Delta r\gtrsim h.
    \end{cases}
\end{equation}

We conclude that during quiescence --- when $\dot{m}\ll \dot{m}_{\rm ph}$ --- the companion star's photosphere underfills its Roche lobe by several $h \sim (T_{\rm eff}/T_{\rm c})r\sim 3\times 10^{-4} r$, where $T_{\rm c}\sim Gm\mu/(kr)\sim 10^7\textrm{ K}$ is the star's central temperature (i.e. the Roche lobe is offset by $\Delta r_{\rm L}\equiv -\Delta r\sim 10^{-3} r$). The exact value of $\Delta r_{\rm L}$ depends on the star's structure exterior to the photosphere, potentially reaching the chromosphere, where the star's atmosphere heats up beyond $T_{\rm eff}$. 

\section{Nova eruption}\label{sec:eruption}

In this section we calculate the response of the donor star to the nova's irradiation under the simplifying assumption of spherical symmetry. However, as the eruption subsides, an accretion disc re-forms around the white dwarf --- potentially shielding the L1 point and the surrounding nozzle from direct irradiation. None the less, atmospheric circulation can convey hot material from illuminated latitudes to L1. These multidimensional effects have been studied, usually in the context of dwarf novae, with an uncertain conclusion \citep[e.g.][]{Sarna1990,Smak2004a,Smak2004b,VialletHameury2007,VialletHameury2008,Cambier2015}, and they are beyond the scope of this paper. The accretion rates that we calculate below are therefore an upper limit and they might be reduced due to the shielding of L1.

\subsection{Heat penetration}\label{sec:wave}

When a nova erupts, it irradiates the companion star and sets an outer boundary temperature $T_{\rm irr}>T_{\rm eff}$. This hot layer gradually penetrates deeper (in a Lagrangian  sense) by diffusion, heating a progressively larger mass to a temperature $\approx T_{\rm irr}$ \citep{Kovetz88}. The time it takes to heat a layer with a mass $\Delta m\ll m$ by diffusion to a temperature $T\gtrsim 10^4\textrm{ K}$ is
\begin{equation}\label{eq:t_diff}
    t\sim\frac{\Delta m k T/\mu}{4\upi r^2\sigma T_{\rm irr}^4/\tau}\sim\frac{\kappa k\Sigma^2T}{\mu\sigma T_{\rm irr}^4}\propto \frac{m\Sigma^3}{r^2T_{\rm irr}^4T^{7/2}},
\end{equation}
where $\tau\gg 1$ is the layer's optical depth and $\sigma$ denotes the Stefan--Boltzmann constant. We assume here that the irradiation is deposited close enough to the photosphere such that the atmosphere is heated to $\sim T_{\rm irr}$. In this case, the heating flux is $\sim \sigma {\rm d}T^4/{\rm d}\tau\sim\sigma T_{\rm irr}^4/\tau$. This assumption is justified self-consistently in Section \ref{sec:numerical_wave}. In the second similarity in equation \eqref{eq:t_diff} we used $\tau\sim \kappa\Sigma$, where $\Sigma\equiv\Delta m/(4\upi r^2)$ is the mass column density. In the final proportionality, we assumed Kramers' opacity $\kappa\propto\rho T^{-7/2}$ \citep[e.g.][]{Hansen2004,Kippenhahn2012}, and used $\Sigma\sim\rho h$, where the heated layer expands by a scale height $h=kr^2T/(Gm\mu)$.
Equation \eqref{eq:t_diff} indicates that the heating time to $T_{\rm irr}$ is actually dominated by the time spent at lower temperatures $T<T_{\rm irr}$, such that the penetration of the heat with time scales as
\begin{equation}\label{eq:delta_m_anal}
    \Delta m_{\rm heat}(t)=4\upi r^2\Sigma(t)\propto r^{8/3}m^{-1/3}T_{\rm irr}^{4/3}t^{1/3}.
\end{equation} 

\subsubsection{Numerical computation}\label{sec:numerical_wave}

\begin{figure}
\includegraphics[width=\columnwidth]{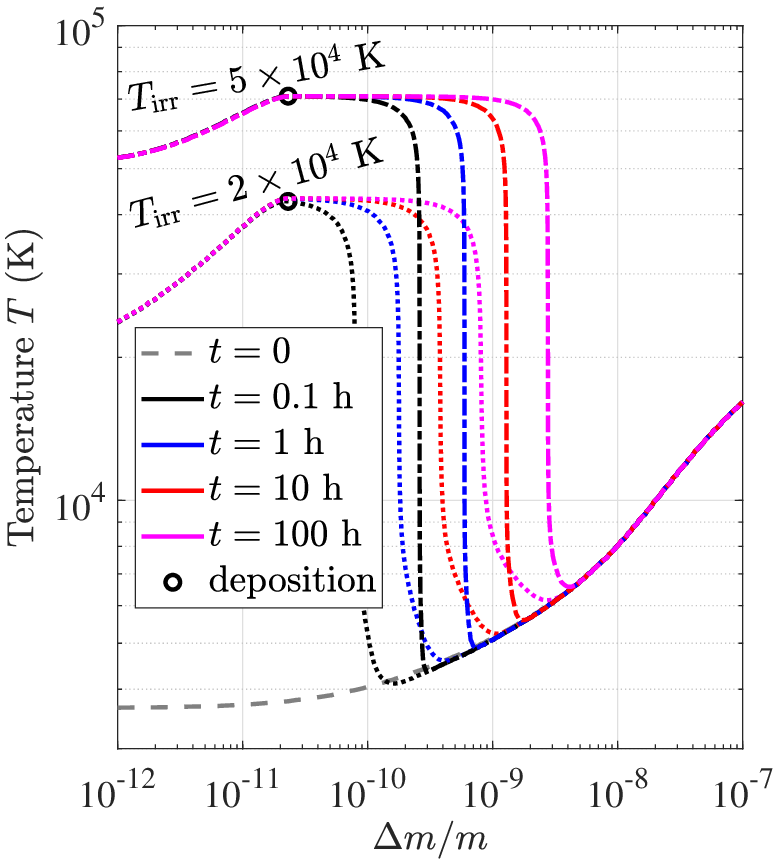}
\caption{Heat penetration into an $m=0.25$ ${\rm M}_\odot$ companion ($r\approx 0.25$ ${\rm R}_\odot$), calculated using \textsc{mesa} for two nova irradiation temperatures $T_{\rm irr}=2\times 10^4$ K (dotted lines) and $T_{\rm irr}=5\times 10^4$ K (dot--dashed lines). The heating begins at $t=0$ (dashed grey line) and penetrates progressively deeper into the companion $\Delta m_{\rm heat}\propto t^{1/3}$. Circles mark the depth at which the irradiation is deposited (this is an upper limit; see text).}
\label{fig:wave}
\end{figure}

In Fig. \ref{fig:wave} we present numerical calculations of the heat penetration into a 0.25 ${\rm M}_\odot$ companion, performed using the stellar evolution code \textsc{mesa} \citep{Paxton2011,Paxton2013,Paxton2015,Paxton2018,Paxton2019}. We implement the irradiation using the $F_\star$--$\Sigma_\star$ method \citep{Paxton2013}: $F_\star=4\sigma T_{\rm irr}^4$ is the incoming flux, which we parametrize using the induced effective temperature $T_{\rm irr}$ (assuming efficient redistribution between the two hemispheres), and $\Sigma_\star=3\textrm{ g cm}^{-2}$ is the mass column density for electron scattering, which sets an upper limit on the irradiation depth. Since $\Sigma_\star$ is below the photosphere, the atmosphere is heated to a somewhat higher temperature than $T_{\rm irr}$, as seen in Fig. \ref{fig:wave}. In reality, the white dwarf's EUV photons are deposited between the photosphere and our chosen $\Sigma_\star$, such that we overestimate the temperature of the heated layer by less than a factor of $\approx 2$. This effect is significant only for low $T_{\rm irr}$, when the temperature at the Roche lobe cannot be accurately determined anyway (see Fig. \ref{fig:profile}).
As seen in Fig. \ref{fig:wave}, the heat penetrates by diffusion into $\Sigma>\Sigma_\star$ in less than 0.1 h. For numerical convenience, we linearly increase the flux from zero to $F_\star$ over a rise time of $t=100$ s, which is much shorter than any relevant time-scale in the problem.
Fig. \ref{fig:wave} agrees well with the analytical scaling found in equation \eqref{eq:delta_m_anal}.

\subsection{Boosted Roche-lobe overflow}\label{sec:rlo}

During a nova eruption, the heated outer layers of the companion expand such that it overflows its Roche lobe more than in quiescence --- significantly amplifying the mass transfer through L1. Unlike quiescence, a nova outburst is too short for magnetic braking or gravitational waves to adjust the orbit and the size of the Roche lobe --- $m/\dot{m}$ is now decoupled from these angular momentum loss mechanisms (we discuss faster orbital change below). 

We calculate the mass transfer rate $\dot{m}$ similarly to Section \ref{sec:quiescence}:
\begin{equation}\label{eq:m_loss_heat}
    \dot{m}\equiv\dot{m}_{\rm loss}\approx\frac{2\upi F}{\sqrt{\rm e}}\rho c_{\rm s}r\Delta r=\frac{F}{2\sqrt{\rm e}}\frac{c_{\rm s}}{r}\Delta m_{\rm heat}=\frac{3F}{2\sqrt{\rm e}}\frac{c_{\rm s}t}{r}\dot{m}_{\rm heat},
\end{equation}
where $\Delta r\sim h(T_{\rm irr})= kr^2T_{\rm irr}/(Gm\mu)$ is the inflation beyond the Roche lobe (assuming hydrostatic equilibrium, which is justified below). Since in quiescence $\Delta r\approx -3h(T_{\rm eff})$ (see Section \ref{sec:quiescence}), and since the heated layers expand by a larger scale height $h(T_{\rm irr}\gg T_{\rm eff})$ once the nova erupts, we approximately identify the Roche lobe with the companion's radius before eruption, i.e. $r(t=0)$; see details below. The density $\rho$ at the bottom of the heated (and inflated) zone is given by the mass that the heat has penetrated by time $t$, as calculated in Section \ref{sec:wave}:
$4\upi r^2\Delta r\rho=\Delta m_{\rm heat}(t)\propto t^{1/3}$. The speed of sound is $c_{\rm s}\sim (kT_{\rm irr}/\mu)^{1/2}$.

Equation \eqref{eq:m_loss_heat} is applicable only for times $t\leq t_{\rm max}\sim r/c_{\rm s}$, which is determined by $\dot{m}_{\rm loss}=\dot{m}_{\rm heat}$ (we omit a coefficient of $2\sqrt{\rm e}/(3F)\approx 0.7$ which is considered later). The outer layers of the companion, which are being heated by the irradiation, are the same layers that are being lost through Roche-lobe overflow. As the heated mass is removed, new cool layers are exposed to the heating, requiring $\dot{m}_{\rm loss}\leq\dot{m}_{\rm heat}\equiv{\rm d}\Delta m_{\rm heat}(t)/{\rm d}t$ for a consistent solution \citep{HarpazRappaport91}. In other words, up to a time $t_{\rm max}$, only a fraction of the heated mass has been removed ($\Delta m_{\rm loss}\propto t^{4/3}<\Delta m_{\rm heat}\propto t^{1/3}$). At later times $t>t_{\rm max}$, the heating and mass loss reach a steady state in which a layer $\Delta m_{\rm heat}(t_{\rm max})$ is being both heated and lost on a time-scale $t_{\rm max}$. We conclude that the mass transfer rate $\dot{m}$ saturates at
\begin{equation}\label{eq:loss_heat_prop}
    \dot{m}_{\rm loss}\sim\dot{m}_{\rm heat}\sim\frac{\Delta m_{\rm heat}(t_{\rm max})}{t_{\rm max}}\propto r^2m^{-1/3}T_{\rm irr}^{5/3}\propto m^{1.3}T_{\rm irr}^{5/3}
\end{equation}
with $\Delta m_{\rm heat}(t)$ calculated in equation \eqref{eq:delta_m_anal}.
We note that while our one-dimensional hydrostatic \textsc{mesa} calculation of $\Delta m_{\rm heat}(t)$ is inadequate for time-scales much shorter than the sound crossing time $r/c_{\rm s}$, it should be accurate to within order-unity corrections for $t=t_{\rm max}\sim r/c_{\rm s}$ itself (which allows enough time for material to flow to L1; see however the discussion in the beginning of Section \ref{sec:eruption}). For the final proportionality in equation \eqref{eq:loss_heat_prop} we assumed $r\propto m^{0.8}$ \citep[e.g.][]{Kippenhahn2012}.

\begin{figure}
\includegraphics[width=\columnwidth]{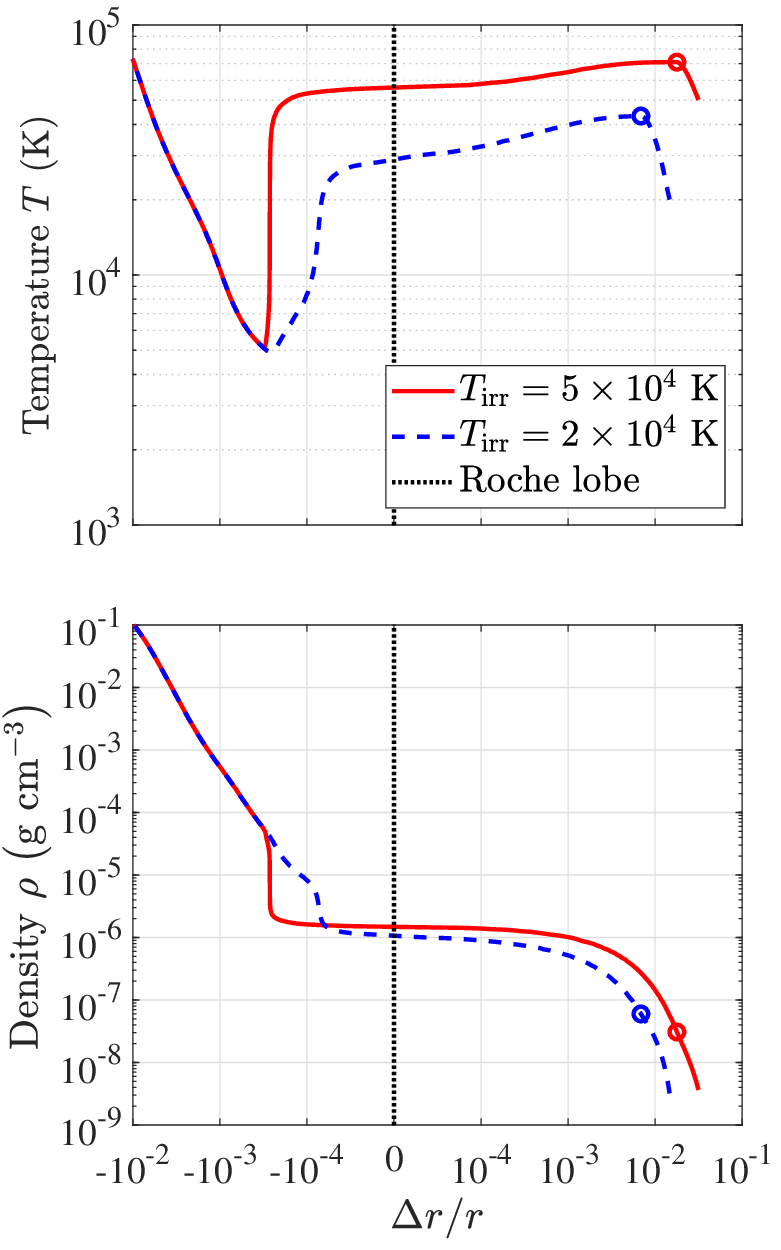}
\caption{Temperature (top panel) and density (bottom panel) profiles of irradiated 0.25 ${\rm M}_\odot$ ($r\approx 0.25$ ${\rm R}_\odot$) companions at the saturation time $t=t_{\rm max}=r/c_{\rm s}\propto T_{\rm irr}^{-1/2}$ for irradiation temperatures $T_{\rm irr}=2\times 10^4$ K (dashed blue lines, $t=3.8$ h) and $T_{\rm irr}=5\times 10^4$ K (solid red lines, $t=2.4$ h). The radial coordinate $\Delta r$ is measured relative to the Roche lobe $r_{\rm L}$ (dotted black lines), which we approximately identify here with the companion's radius at $t=0$ (this is a slight underestimate because in quiescence $\Delta r_{\rm L}/r\sim 10^{-3}$; we refine the calculation in Section \ref{sec:offset}). The horizontal axis is logarithmic on both sides of the Roche lobe, with the $|\Delta r|/r<10^{-5}$ region excised. As in Fig. \ref{fig:wave}, the irradiation deposition depth is marked with circles.}
\label{fig:profile}
\end{figure}

In Fig. \ref{fig:profile} we plot profiles of irradiated companions at the saturation time $t=t_{\rm max}(T_{\rm irr})$ --- these depict the steady-state structure of the companion, from which the mass transfer rate $\dot{m}$ is determined. The figure demonstrates that the companion star is inflated by $\Delta r/r\sim h/r\sim T_{\rm irr}/T_{\rm c}\sim 10^{-3}-10^{-2}$. The bottom panel shows that the density of the inflated region $\rho$ is a weak function of $T_{\rm irr}$: a higher $T_{\rm irr}$ heats a larger mass $\Delta m_{\rm heat}$, but this mass is spread over a larger $h\propto T_{\rm irr}$:
\begin{equation}\label{eq:rho_T_power}
    \rho\sim\frac{\Delta m_{\rm heat}}{4\upi r^2h}\propto\frac{T_{\rm irr}^{4/3}t_{\rm max}^{1/3}}{T_{\rm irr}}\propto T_{\rm irr}^{1/6},
\end{equation}
 where we substituted $\Delta m_{\rm heat}$ from equation \eqref{eq:delta_m_anal} and $t_{\rm max}\sim r/c_{\rm s}$.

The temperature and density profiles in Fig. \ref{fig:profile} are approximately flat in the $-10^{-4}\lesssim\Delta r/r\lesssim h/r\sim 10^{-3}-10^{-2}$ region, where $\Delta r$ is measured relative to the radius of a non-irradiated companion, i.e. $r(t=0)$. 
The extent of the flat region with $\Delta r<0$ is determined by heating and expansion at the moderately rising precursor just below the sharp heating front (e.g. $T\lesssim 10^4\textrm{ K}$ on the rightmost dotted line in Fig. \ref{fig:wave}).
We may therefore approximately identify $r_{\rm L}=r(0)$, when the Roche lobe actually exceeds the photosphere by $\Delta r_{\rm L}/r\sim 10^{-3}$ (Section \ref{sec:quiescence}); this approximation holds for the high $T_{\rm irr}\approx 5\times 10^4\textrm{ K}$ at the peak of a nova eruption, but breaks down as the white dwarf cools. In Section \ref{sec:offset} we repeat the calculation with $\Delta r_{\rm L}/r\sim 10^{-3}$, which is relevant for lower $T_{\rm irr}$, when the eruption subsides.  

Mass ejection from the white dwarf during the nova can expand the orbit of the binary and therefore the companion's Roche lobe by a fraction $\sim 10^{-5}$ \citep[of order the lost mass fraction; see][]{Shara86}. While this small effect can be neglected in the calculation above, some (many decades old) novae have been recently claimed to {\it shrink} the binary orbit (and consequently the Roche lobe) by as much as $\sim 10^{-4}-10^{-3}$ \citep{Salazar2017,Schaefer2020}, which might be enough to modify our $\dot{m}$ estimate. Since we lack a clear picture of how the binary orbit changes during the nova \citep[see][for a recent review]{Chomiuk2020} we ignore this possibility and determine $\dot{m}$ using $r_{\rm L}(t=0)$. While we assume that changes to $r_{\rm L}$ during an eruption can be neglected when calculating a single nova cycle, these changes can accumulate over multiple cycles and affect the long term evolution of CVs; we briefly discuss this in Section \ref{sec:long_term}. 

\section{Mass transfer rates}\label{sec:rates}

\begin{figure}
\includegraphics[width=\columnwidth]{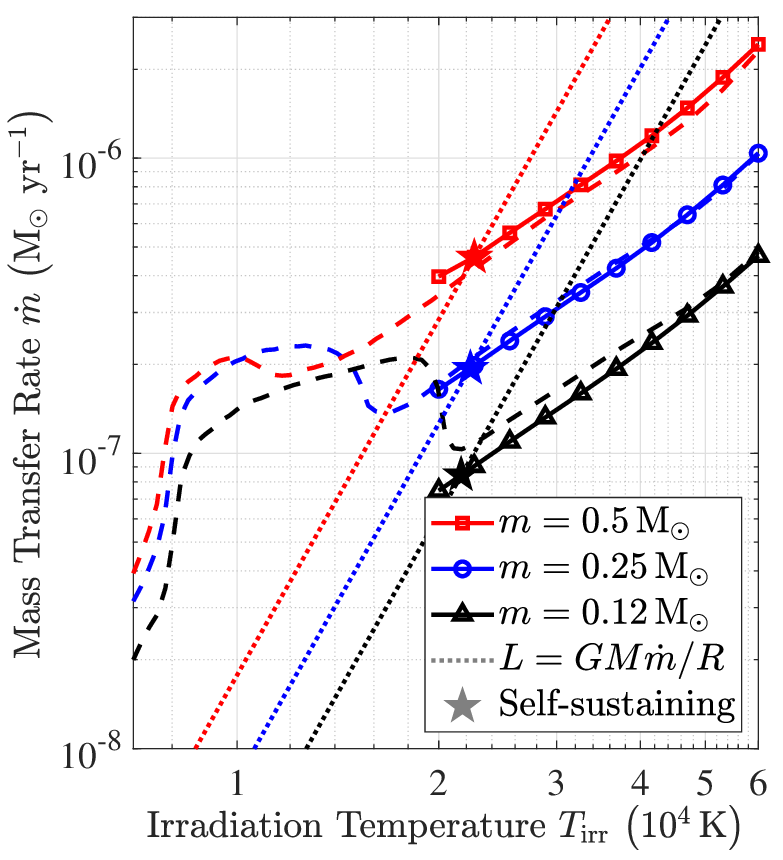}
\caption{Mass transfer rates calculated from \textsc{mesa} using two methods: directly, $\dot{m}=\rho c_{\rm s}r\Delta r$, using $\rho$, $c_{\rm s}$, and $\Delta r$ at the Roche lobe (dashed lines), and indirectly, using the heating rate $\dot{m}=\Delta m_{\rm heat}(t_{\rm max})/t_{\rm max}$ (solid lines with markers). The two methods are consistent for $T_{\rm irr}\gtrsim 2\times 10^4\textrm{ K}$ and agree with our analytical scaling $\dot{m}\propto m^{1.3}T_{\rm irr}^{5/3}$ (equation \ref{eq:loss_heat_prop}). $T_{\rm irr}=5.4\times 10^4\textrm{ K}$ corresponds to a nova luminosity of $L=16\upi a(M,m)^2 \sigma T_{\rm irr}^4=10^{38}\textrm{ erg s}^{-1}$ for an $m=0.25$ ${\rm M}_\odot$ companion orbiting an $M=1$ ${\rm M}_\odot$ white dwarf (using \citealt{Eggleton83} to calculate the binary separation $a$). The dotted lines mark $\dot{m}$ that is required to sustain an accretion luminosity $L=GM\dot{m}/R= 16\upi a(m)^2\sigma T_{\rm irr}^4$ from a solar mass white dwarf. The mass transfer becomes self-sustaining at the intersection of the lines, marked with a star ($\equiv T_{\rm irr}^0\approx 2\times 10^4\textrm{ K}$, with $\dot{m}\sim 10^{-7}\msunyr$).}
\label{fig:rates}
\end{figure}

In Fig. \ref{fig:rates} we calculate the mass transfer rates $\dot{m}$ from \textsc{mesa} profiles at $t=t_{\rm max}= r/c_{\rm s}$ using two methods:
\begin{enumerate}
    \item {\it Heated layer:} $\dot{m}=\Delta m_{\rm heat}/t_{\rm max}$, where $\Delta m_{\rm heat}$ is the mass that is heated above $T_{\rm irr}/2$. The heating front is very sharp, such that we are not sensitive to the exact threshold; see Fig. \ref{fig:wave}. Note that by omitting order-unity coefficients here (for the sake of simplicity), we overestimate $\dot{m}$ by an overall factor of $3^{1/3}(2\sqrt{\rm e}/F)^{2/3}\approx 2.4$, which is not crucial considering the other simplifications and uncertainties of the model and of the observations.
    \item {\it Direct:} $\dot{m}=\rho c_{\rm s}r\Delta r$, with $\rho$ and $c_{\rm s}$ evaluated at $r(t=0)$ and with $\Delta r=r(t_{\rm max})-r(0)$. That is, we identify the Roche lobe with the companion's quiescent radius (see Fig. \ref{fig:profile}).
    Here too we omit order-unity coefficients in order to match the normalization of both methods (see Fig. \ref{fig:rates}).
\end{enumerate}
In light of the discussion in Section \ref{sec:rlo}, the two methods should be equivalent.
This is confirmed by Fig. \ref{fig:rates}, which also agrees with the analytical scaling found in equation \eqref{eq:loss_heat_prop}: $\dot{m}\propto m^{1.3}T^{5/3}$. Below $T_{\rm irr}\lesssim 2\times 10^4\textrm{ K}$ many of our model's assumptions break down. Specifically, the opacity can no longer be approximated by Kramers' law, and $T_{\rm irr}$ becomes comparable to the donor's quiescent temperature, smearing the heating front. The heating at the precursor below the front (Fig. \ref{fig:wave}) leads to its expansion, such that $r_{\rm L}$ no longer falls inside the region with flat $T$ and $\rho$ (Fig. \ref{fig:profile}).
Both our numerical methods and the analytical solution lose accuracy at these low irradiation temperatures. As we show below, we are interested mainly in $T_{\rm irr}\gtrsim 2\times 10^4\textrm{ K}$ anyway; we provide the direct solution for $T_{\rm irr}<2\times 10^4\textrm{ K}$ in Fig. \ref{fig:rates} only as a rough approximation.

Following a short thermonuclear runaway at the onset of a nova eruption, the white dwarf's luminosity reaches $\sim 10^{38}\textrm{ erg s}^{-1}$ --- similar to the Eddington luminosity --- equivalent to $T_{\rm irr}\approx 5\times 10^4\textrm{ K}$ on the companion's surface \citep{Prialnik86,Gehrz98,Chomiuk2020}. According to Fig. \ref{fig:rates}, the mass transfer rate reaches $\sim 10^{-6}\msunyr$ --- about a thousand times the quiescent value set by gravitational waves and magnetic braking \citep{Patterson84}. The white dwarf then cools down, and over a time-scale of roughly a year its luminosity drops by two orders of magnitude, such that $T_{\rm irr}\approx 2\times 10^4\textrm{ K}$ \citep{Prialnik86,Hillman2014}.
It is clear from Section \ref{sec:rlo} that $\dot{m}$ adjusts to changes in $T_{\rm irr}$ on a time-scale of $r/c_{\rm s}$, which is measured in hours --- practically instantaneously compared to the decline of $T_{\rm irr}$.

\subsection{Comparison with previous calculations}\label{sec:previous}

\begin{figure}
\includegraphics[width=\columnwidth]{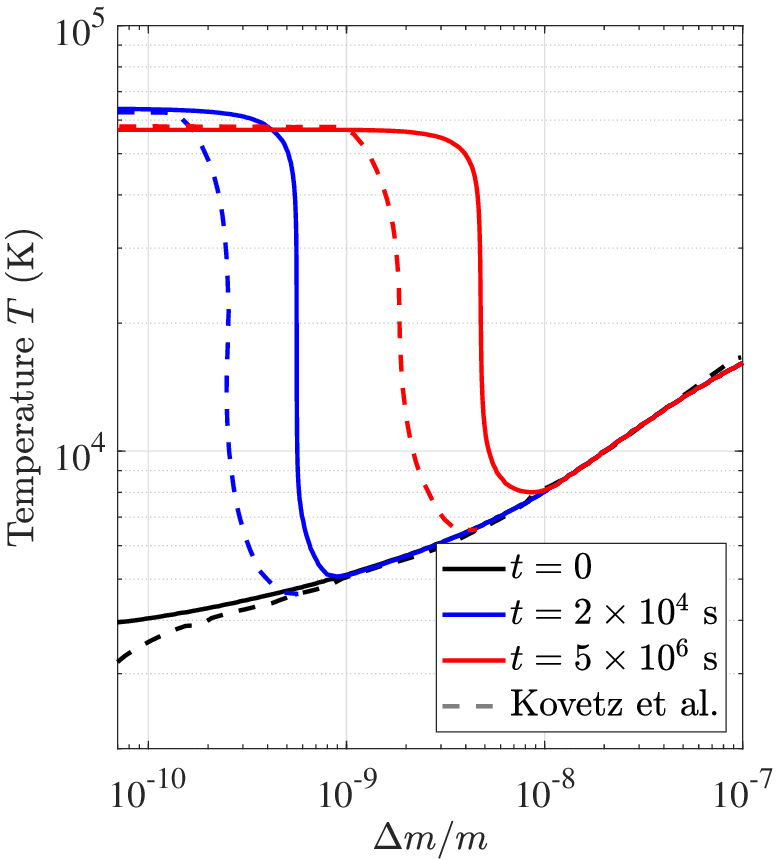}
\caption{Heat penetration into an $m=0.25$ ${\rm M}_\odot$ companion: comparing our model (solid lines) with \citet[][dashed lines]{Kovetz88} for the same irradiation. The irradiating flux rises linearly from zero to its maximal value over a time $t=2\times 10^4\textrm{ s}$ and it gradually decays starting from a time $t=3\times 10^6\textrm{ s}$, as given by equation (1) of \citet{Kovetz88}. We speculate that the factor of $\approx 2$ difference in the penetration depth is the result of using modern opacity tables in \textsc{mesa} \citep{Paxton2011}. In any case, such a difference does not significantly affect our conclusions.}
\label{fig:kovetz}
\end{figure}

\citet{Kovetz88} conducted similar calculations of the response of a red dwarf companion to a nova eruption. More recently, \citet{Hillman2020} evolved CVs for several gigayears and through numerous nova eruptions, incorporating the effects of nova irradiation as calculated by \citet{Kovetz88}. The heat fronts in our calculations and in \citet{Kovetz88} have similar shapes and they scale similarly with time, as demonstrated in Fig. \ref{fig:kovetz}. \citet{Hameury1988} find a similar heat front as well, which advances roughly as $t^{1/2}$ (compared to our $t^{1/3}$) in the context of soft X-ray transients (see their fig. 1).

While the penetration of heat into the companion's atmosphere is calculated similarly here and in \citet{Kovetz88}, the inference of the mass transfer rate $\dot{m}$ is conceptually different.
\citet{Kovetz88} assumed that the density $\rho$ at L1 remains similar to the photospheric density of the quiescent star $\rho_{\rm ph}$, up to an exponential factor of order unity (because $\Delta r\sim h)$. 
The resulting mass transfer rate in their case is
$\dot{m}\sim\rho_{\rm ph} r c_{\rm s}(T_{\rm irr}) h(T_{\rm irr})\propto T_{\rm irr}^{3/2}$. 
We, on the other hand, find $\rho\propto T_{\rm irr}^{1/6}$ by consistently calculating the mass and expansion of the heated zone (equation \ref{eq:rho_T_power}) --- yielding $\dot{m}\propto \rho rc_{\rm s}h\propto T_{\rm irr}^{5/3}$. The omission of this weak variation of $\rho$ with $T_{\rm irr}$ by \citet{Kovetz88} is partially compensated by the exponential factor in their model. In addition, equation \eqref{eq:loss_heat_prop} indicates that for the same $T_{\rm irr}$, $\dot{m}\propto m^{1.3}$, much steeper than the mass dependence found by \citet[][their table 3]{Kovetz88}.

More important than correcting these power laws (the dependence on the temperature hardly changes) is normalizing them. \citet{Kovetz88} arbitrarily calibrate $\dot{m}=10^{-9}\msunyr$ for a 0.5 ${\rm M}_\odot$ companion 100 yr after an eruption.
\citet{Hillman2020} suggest that the mass transfer rate is enhanced by a factor of $(T_{\rm irr}/T_{\rm eff})^{3/2}\sim 10^2$ relative to quiescence. In our model, on the other hand, the quiescent $\dot{m}$ is unrelated to $T_{\rm eff}$: as explained in Section \ref{sec:quiescence}, the Roche-lobe overfilling length $\Delta r(\dot{m})$ adjusts to accommodate the mass transfer rate, which is set by magnetic braking and gravitational waves. Consequently, there is no `enhancement factor': $\dot{m}$ during and immediately after a nova eruption is a function of $T_{\rm irr}$ and $m$ alone, regardless of the quiescent rate --- the normalization is given in Fig. \ref{fig:rates}. The bottom line is that we find mass transfer rates that are more than an order of magnitude higher than \citet{Kovetz88} and \citet{Hillman2020}.

\subsection{Bootstrapping (self-sustaining accretion)}\label{sec:boots}

Fig. \ref{fig:rates} shows that as the white dwarf cools down, and $\dot{m}$ decreases, the mass transfer onto the white dwarf becomes self-sustaining. At the intersection of the solid and dotted lines, $\dot{m}$ is just enough to lift itself out through L1 using the accretion luminosity $L=GM\dot{m}/R$ ($M$ and $R$ are the white dwarf's mass and radius, respectively). 
We find that the self-sustaining rate is $\dot{m}\sim 10^{-7}\msunyr$ (equivalently $L\sim 10^{36}\textrm{ erg s}^{-1}$), providing a possible explanation for the puzzling high accretion rates of the recurrent novae T Pyx and IM Nor during quiescence \citep[][]{Patterson2017,Patterson2020,Godon2018}. \citet{Knigge2000} proposed a different self-sustaining mechanism for T Pyx: an irradiation-driven wind carries away angular momentum from the binary, driving the companion to overflow its Roche lobe.
\citet{Knigge2000} rely on stable nuclear burning to power the evaporation (see Section \ref{sec:stable}), whereas our mechanism can explain T Pyx with the accretion luminosity alone (which is weaker by a factor of $\approx 30$). 
Moreover, our mechanism remains valid even if the binary conserves its total mass and angular momentum.  

\subsubsection{Roche-lobe offset}\label{sec:offset}

This high quiescent accretion rate might not follow all novae, as also suggested observationally \citep{Patterson2017,Patterson2020}, because it depends critically on the size of the donor's Roche lobe. In Sections \ref{sec:eruption} and \ref{sec:rates} we have so far assumed that the companion's Roche lobe $r_{\rm L}$ coincides with its un-inflated radius $r(t=0)$. In Section \ref{sec:quiescence}, however, we found that angular momentum loss by magnetic braking and gravitational waves dictates an offset of $\Delta r_{\rm L}\equiv r_{\rm L}-r(0)\sim 10^{-3} r$. We repeat the calculation in Section \ref{sec:rlo} with $\Delta r_{\rm L}>0$ and find that 
\begin{equation}\label{eq:Delta_m_dot}
    \dot{m}\sim \rho c_{\rm s}r(h-\Delta r_{\rm L})\sim \Delta m_{\rm heat}(t) \frac{c_{\rm s}}{r}\left(1-\frac{\Delta r_{\rm L}}{h}\right).
\end{equation}
Equation \eqref{eq:Delta_m_dot} saturates at $t_{\rm max}=(r/c_{\rm s})(1-\Delta r_{\rm L}/h)^{-1}>r/c_{\rm s}$ and the mass transfer rate scales as
\begin{equation}\label{eq:Delta_m_dot_scaling}
    \dot{m}\sim\frac{\Delta m_{\rm heat}(t_{\rm max})}{t_{\rm max}}\propto T_{\rm irr}^{5/3}\left[1-\frac{\Delta r_{\rm L}}{h(T_{\rm irr})}\right]^{2/3}.
\end{equation}
At the peak of a nova eruption, the offset $\Delta r_{\rm L}$ hardly changes $\dot{m}$ because $\Delta r_{\rm L}/h(T_{\rm irr})\sim 0.1$. During the cooler self-sustaining accretion phase that follows, on the other hand, $h$ becomes comparable to $\Delta r_{\rm L}$ --- significantly lowering $\dot{m}$ and with it $h\propto T_{\rm irr}\propto L^{1/4}\propto \dot{m}^{1/4}$.
Using equation \eqref{eq:Delta_m_dot_scaling}, the self-sustaining irradiation temperature $T_{\rm irr}(\Delta r_{\rm L})\propto \dot{m}^{1/4}$ scales as $T_{\rm irr}\propto (1-\Delta r_{\rm L}/h)^{2/7}$. Explicitly,
\begin{equation}\label{eq:Delta_Tirr}
\frac{T_{\rm irr}}{T_{\rm irr}^0}=\left[1-\frac{\Delta r_{\rm L}}{h(T_{\rm irr}^0)}\frac{T_{\rm irr}^0}{T_{\rm irr}}\right]^{2/7},    
\end{equation}
where $T_{\rm irr}^0\approx 2\times 10^4\textrm{ K}$ is the solution for $\Delta r_{\rm L}=0$ (i.e. the star markers in Fig. \ref{fig:rates}) and $h(T_{\rm irr}^0)\sim 10^{-3} r$ is the corresponding scale height.
We solve equation \eqref{eq:Delta_Tirr} for different values of $\Delta r_{\rm L}$ in Fig. \ref{fig:delta_rl} (the stability of the solution is discussed in Appendix \ref{sec:stability}). As $\Delta r_{\rm L}$ increases, the self-sustaining $T_{\rm irr}\propto\dot{m}^{1/4}$ gradually drops; there is no solution for $\Delta r_{\rm L}/h(T_{\rm irr}^0)> 0.51$ --- the donor recedes within its Roche lobe and the mass transfer rate drops exponentially (see Section \ref{sec:quiescence}). 

\begin{figure}
\includegraphics[width=\columnwidth]{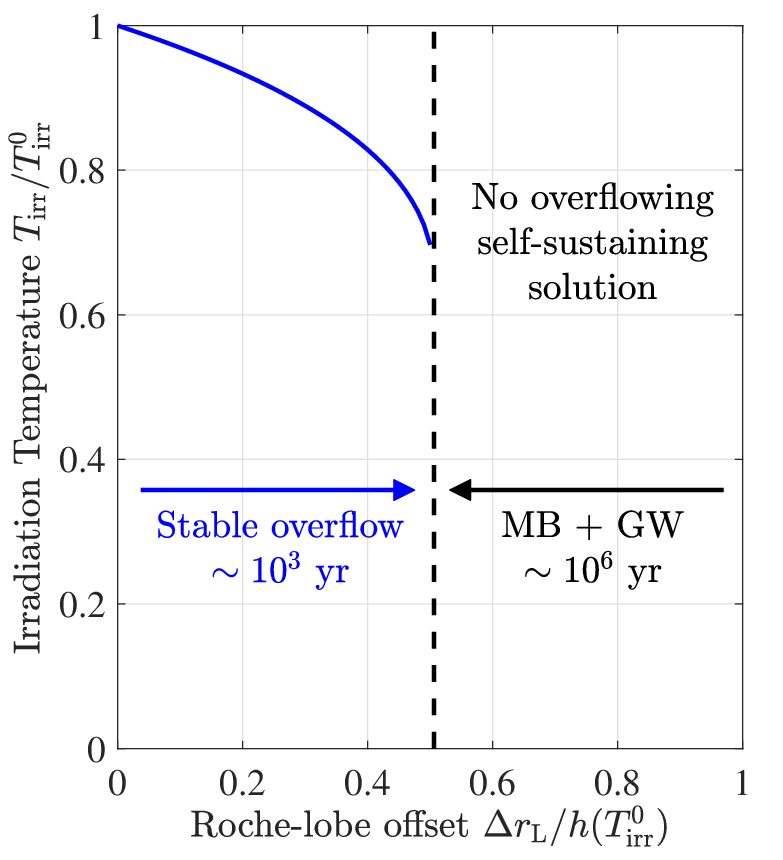}
\caption{The self-sustaining (bootstrapping) irradiation temperature $T_{\rm irr}\propto\dot{m}^{1/4}$ 
as a function of $\Delta r_{\rm L}\equiv r_{\rm L}-r(0)$, the difference between the donor's Roche lobe $r_{\rm L}$ and its non-irradiated radius $r(0)$. $T_{\rm irr}^0\equiv T_{\rm irr}(\Delta r_{\rm L}=0)\approx 2\times 10^4\textrm{ K}$ is calculated in Fig. \ref{fig:rates}, and $h(T_{\rm irr}^0)\sim 10^{-3}r$ is the corresponding scale height. The curve is given by equation \eqref{eq:Delta_Tirr}, which has no solution for $\Delta r_{\rm L}/h(T_{\rm irr}^0)> 0.51$, where the donor recedes within its Roche lobe and $\dot{m}$ drops steeply. Stable Roche-lobe overflow increases $\Delta r_{\rm L}$ by the critical $\sim 10^{-3}r$ over $t\sim 10^3\textrm{ yr}$, whereas angular momentum loss by magnetic braking (MB) and gravitational waves (GW) decreases $\Delta r_{\rm L}$ by a similar amount over $t\sim 10^6\textrm{ yr}$ --- bootstrapping with a quiescent $\dot{m}\sim 10^{-7}\msunyr$ is a short lived phase (though the exact long term evolution of $\Delta r_{\rm L}$ is uncertain; see Section \ref{sec:long_term}).}
\label{fig:delta_rl}
\end{figure}

\subsubsection{Long term evolution}\label{sec:long_term}

Why do some novae remain locked in the self-sustaining state whereas others decline to lower quiescent accretion rates? The Roche-lobe's size before and immediately after an eruption may hold the key. 
The exact value of $\Delta r_{\rm L}$ before the eruption depends on the donor's density structure at several scale heights above the photosphere, as indicated by equation \eqref{eq:t_loss_general}. At such heights, the profile might deviate from our isothermal approximation due to heating by e.g. magnetic fields \citep{AvrettLoeser2008}, making it hard to establish in which of the two regimes of Fig. \ref{fig:delta_rl} $\Delta r_{\rm L}$ falls. 

Even if the donor initially satisfies $\Delta r_{\rm L}< 0.51 h(T_{\rm irr}^0)\sim 10^{-3}r$, the high self-sustaining accretion rate may not last indefinitely because the size of the Roche lobe changes over time. The mass transfer in CVs is stable --- it increases $\Delta r_{\rm L}$ and drives the companion away from Roche-lobe overflow $(1/r){\rm d}\Delta r_{\rm L}/{\rm d}t\sim \dot{m}/m$ \citep[the orbit expands to conserve angular momentum; see][]{Rappaport82}. At the self-sustaining mass transfer rate of $\dot{m}\sim 10^{-7}\msunyr$, the critical offset of $\Delta r_{\rm L}\sim 10^{-3}r$ is reached within $t\sim 10^3\textrm{ yr}$, ending the bootstrapping phase. 
Loss of angular momentum by magnetic braking and gravitational waves pushes the companion back towards Roche-lobe overflow and decreases $\Delta r_{\rm L}$, but this process unfolds on a longer time-scale: it takes $\sim\textrm{Gyr}$ to shrink the orbit by half, and hence $t\sim 10^6\textrm{ yr}$ to decrease $\Delta r_{\rm L}$ by $10^{-3} r$. This asymmetry in time-scales may explain why T Pyx and IM Nor are the exception rather than the rule \citep[most novae decline to much lower accretion rates by a similar time after eruption; see][but also the discussion of recurrence times below]{Patterson2017,Patterson2020}.

The increase of $r_{\rm L}$ and the termination of the bootstrapping phase is analogous to the famous hibernation scenario for novae \citep{Shara86}. In that scenario, mass ejection from the white dwarf during an eruption increases $r_{\rm L}$, which reduces $\dot{m}$ (`hibernation'), but only after the irradiated companion has contracted from its inflated radius. In our case, the increase in $r_{\rm L}$ is gradual and it is driven by $\dot{m}$ rather than by sudden mass ejection in an eruption. Specifically, $\dot{r}_{\rm L}/r_{\rm L}\sim \dot{m}/m$ \citep{Rappaport82}, where $\dot{m}$ itself depends on $r_{\rm L}$ (Fig. \ref{fig:delta_rl}). This means that the growth of $r_{\rm L}$ gradually slows down and eventually stops (and $\dot{m}$ stops dropping) exactly when the increase in $r_{\rm L}$ due to the mass transfer is balanced by its decrease due to angular momentum loss by magnetic braking and gravitational waves --- i.e. the value calculated in Section \ref{sec:quiescence}. As we discuss below, additional changes to $r_{\rm L}$ due to mass ejection during eruptions \citep[as considered by][]{Shara86} complicate this picture and may lead to an even lower $\dot{m}$ (true hibernation, when mass transfer virtually stops until gravitational radiation and magnetic braking restore Roche-lobe contact).

Multiple novae can erupt during both bootstrapping periods (when $\Delta r_{\rm L}$ is small such that the quiescent $\dot{m}\sim 10^{-7}\msunyr$) and dormant times (when $\Delta r_{\rm L}$ is large such that $\dot{m}\sim 10^{-10}-10^{-9}\msunyr$, as set by magnetic braking and gravitational waves). The higher accretion rate during bootstrapping drives novae more frequently \citep{Yaron2005,Wolf2013,Chomiuk2020}, explaining the short recurrence times of T Pyx and IM Nor ($10-100\textrm{ yr}$). Dormant periods last $\sim 10^3$ times longer than bootstrapping, but the nova recurrence time is also longer by a similar factor, producing roughly the same total number of novae. T Pyx and IM Nor represent about 30 per cent of the novae in short-period CVs in the \citet{Patterson2020} sample (shorter than about 3 h); the ratio drops to about 10 per cent when longer period systems are included. A high $\dot{m}$ was measured for T Pyx after both the 1966 and 2011 eruptions \citep[see][]{Patterson2017}. It is also worth mentioning the \citet{SchaeferCollazzi2010} sample of novae that are somewhat over-luminous decades after an eruption (they included T Pyx in this class). While these could also potentially be explained by irradiation, their typical inferred $T_{\rm irr}\approx 8\times 10^3\textrm{ K}$ lie in a region of the parameter space where our model's predictions are much more uncertain (see Fig. \ref{fig:rates}). 

We note that orbital periods and therefore $\Delta r_{\rm L}$ jump during eruptions due to ejection of mass and angular momentum (see discussion in Section \ref{sec:rlo}), which might induce or end a bootstrapping phase. The sign and magnitude of these jumps is uncertain \citep{Shara86,Livio1991,Martin2011,Salazar2017,Schaefer2020,Chomiuk2020}, complicating our analysis of the relative numbers of the two nova populations. As an extreme example, if each nova shrinks the orbit by an average fraction of $10^{-5}-10^{-4}$, then eruptions during bootstrapping can counter the expansion of $\Delta r_{\rm L}$ over the $10-100\textrm{ yr}$ recurrence time between novae, when $m/\dot{m}\sim 10^6\textrm{ yr}$. In this case, the bootstrapping phase may proceed all the way up to the donor's destruction \citep[as suggested by][]{Patterson2017,Patterson2020}, which might be further accelerated by unstable mass transfer \citep[triggered by the angular momentum loss in the novae; see][]{Schreiber2016}. While the 2011 T Pyx eruption actually {\it increased} its orbital period \citep{Patterson2017} --- expediting the end of bootstrapping --- some novae might shrink the orbit enough to keep systems in the high $\dot{m}$ state \citep[see table S2 in][]{Chomiuk2020}.

It is important to stress that in our model some variation in either $\Delta r_{\rm L}$ during quiescence (e.g. due to variations from star to star in the upper atmosphere structure of M dwarf donors) or in the orbital period jumps during novae is required to explain why --- at the same orbital period --- some systems are trapped in a self-sustaining state while others are not. Absent such variation from system to system we would expect all systems of a given orbital period and white dwarf mass to eventually evolve towards one or the other of the two solution regimes in Fig. \ref{fig:delta_rl}. One possibility is that eruptions during bootstrapping change the orbital period differently from `regular' novae because of the higher $\dot{m}$ before eruptions. The higher accretion rate raises the temperature of the accreted layer, implying a somewhat lower hydrogen ignition mass \citep[fig. 8 in][]{Wolf2013}. When the nova erupts, the lighter expelled shell in this case is expected to exert less friction on the binary --- removing less angular momentum \citep{Livio1991,Martin2011}. This systematic difference could result in cyclic transitions between the low-$\dot{m}$ and the high-$\dot{m}$ states. Another possibility is that the orbital period jumps are stochastic --- an atypically large negative jump can trigger a temporary bootstrapping phase. Such jumps have been inferred for some historic nova eruptions \citep{Salazar2017,Schaefer2020}. In the case of T Pyx, \citet{Schaefer2010} find evidence that the current high-$\dot{m}$ phase began circa 1866, triggered by a non-recurrent eruption (that is, with a long recurrence time and a low $\dot{m}$ before 1866). This is consistent with our $\sim 10^3\textrm{ yr}$ estimate for the duration of the bootstrapping phase.

\subsubsection{Hydrogen burning stability}\label{sec:stable}

\begin{figure}
\includegraphics[width=\columnwidth]{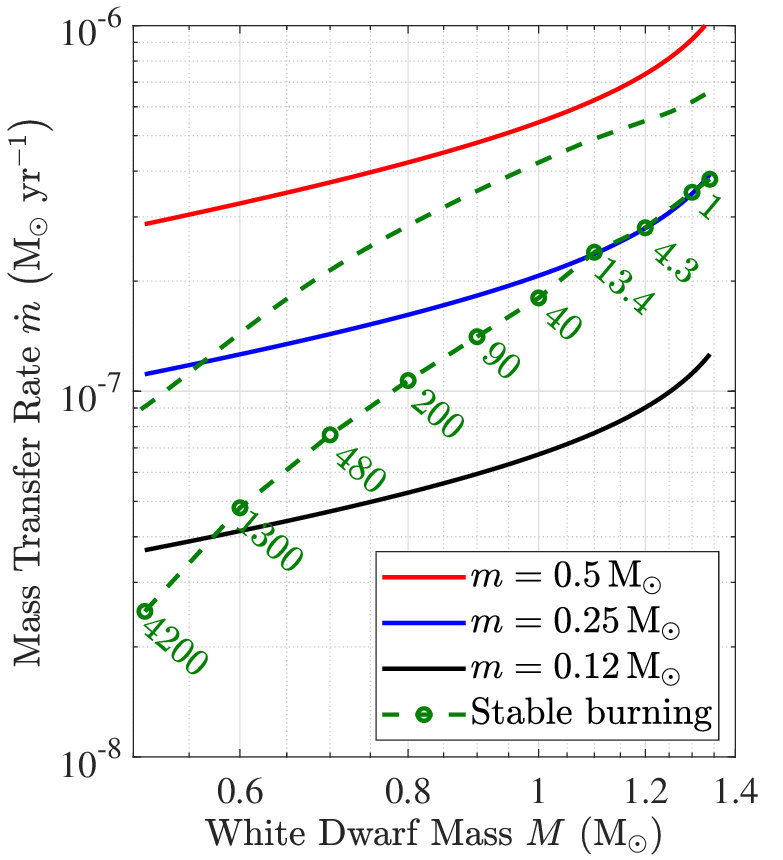}
\caption{Self-sustaining mass transfer rates $\dot{m}$ (assuming $\Delta r_{\rm L}=0$) as a function of the white dwarf's mass $M$ for different companion masses $m$. The rates are calculated by equating $\dot{m}\propto m^{1.3}T_{\rm irr}^{5/3}$ (calibrated using Fig. \ref{fig:rates}) to $\dot{m}=16\upi a^2\sigma T_{\rm irr}^4 R/(GM)$; we take $R(M)$ from \citet{Hansen2004} and calculate $a(M,m)$ using \citet{Eggleton83}. Below the bottom dashed green line \citep[given by][]{Wolf2013}, $\dot{m}$ drives recurrent nova eruptions. Above the line, the white dwarf burns accreted hydrogen stably, powering a supersoft X-ray source. During stable burning, nuclear power increases the irradiation and hence $\dot{m}$. We assume that the mass burning rate saturates at its maximal stable value \citep[the top dashed green line, given by][]{Wolf2013}, and that the rest of $\dot{m}$ is blown away in a wind, as suggested by \citet{Hachisu1996}. The nova recurrence times just below stability are labelled in years \citep[from][]{Wolf2013}. 
}
\label{fig:MWD}
\end{figure}

The high self-sustaining accretion rates that we find $\dot{m}\sim 10^{-7}\msunyr$ are close to the hydrogen burning stability limit \citep{Nomoto2007,ShenBildsten2007,Wolf2013}. If $\dot{m}$ is high enough, then depending on its mass, a white dwarf may stably burn accreted hydrogen instead of erupting as recurrent novae. In Fig. \ref{fig:MWD} we repeat the calculation of the self-sustaining $\dot{m}$ for a range of white dwarf masses $M$ and compare it to the stability limit. Fig. \ref{fig:MWD} indicates that massive white dwarfs with low-mass companions tend to erupt as (recurrent) novae during bootstrapping, whereas lighter white dwarfs with massive companions burn hydrogen stably.
White dwarfs in CVs are typically massive in comparison to their solitary counterparts, with an average $M\approx 0.8\,{\rm M}_\odot$ \citep{Knigge2006,Savoury2011,Zorotovic2011}. For such a typical white dwarf, the critical companion mass above which burning is stable is $m\approx 0.2\,{\rm M}_\odot$, coincidentally close to the mass at the famous CV orbital period gap \citep[$\approx 2-3$ h, see][]{Knigge2011}.
We conclude that short-period CVs (below the gap) like T Pyx (1.8 h) and IM Nor (2.5 h) tend to erupt as recurrent novae when the mass transfer is self-sustaining, whereas long-period CVs (above the gap) can burn hydrogen stably. 

Stable nuclear burning increases the white dwarf's luminosity by a factor of $f\approx 30$ compared to gravitational energy release for the same $\dot{m}$ \citep[e.g.][]{Wolf2013}, leading to stronger irradiation $T_{\rm irr}^4\propto \dot{m}f$. Using our analytical scaling $\dot{m}\propto T_{\rm irr}^{5/3}$, the self-sustaining mass transfer rate is higher by a factor of $\dot{m}\propto f^{5/7}\approx 10$. This suggests a self-sustaining $\dot{m}\sim 10^{-6}\msunyr$, which is sufficiently high such that the white dwarf cannot steadily burn all of the accreted hydrogen \citep[i.e. above the top dashed green line in Fig. \ref{fig:MWD}; see][]{Wolf2013}.
We speculate that the burning rate saturates at the maximal stable value, and the rest of $\dot{m}$ is blown away in a wind \citep{Hachisu1996}.

Stable burning at rates of $\sim 10^{-7}\msunyr$ appears as a prolonged supersoft X-ray phase \citep{KahabkaVanDenHeuvel1997}. Supersoft sources are usually explained by either post-nova burning of a residual hydrogen envelope that is exhausted within several years \citep[for a $0.8\,{\rm M}_\odot$ white dwraf, see][]{Soraisam2016}, or by unstable mass transfer on a thermal time-scale from a massive donor \citep[$m\gtrsim 1.3\,{\rm M}_\odot$ and also heavier than the white dwarf, see][]{vanDenHeuvel92,KahabkaVanDenHeuvel1997}.
Some sources, however, fit neither of these scenarios: the orbital periods of RX J0537.7--7034 (3.5 h) and 1E 0035.4--7230 (4.1 h) are too short for a massive companion, and yet their supersoft emission persists for many decades \citep{vanTeeselingKing98,Greiner2000,King2001}. These sources, with periods above the gap, are naturally explained by our model as the result of self-sustaining mass transfer induced by irradiation. \citet{vanTeeselingKing98} proposed a similar idea, but in their case an irradiation-driven wind pushes the companion towards Roche-lobe overflow by carrying away angular momentum --- our inflation mechanism applies even when mass and angular momentum are conserved \citep[see also][who applied the \citealt{vanTeeselingKing98} mechanism to T Pyx]{Knigge2000}. \citet{Shen2009} found that helium core white dwarfs (with $M<0.5\,{\rm M}_\odot$) enter prolonged supersoft phases after novae, potentially explaining RX J0537.7--7034 and 1E 0035.4--7230. Our scenario, on the other hand, is also valid for regular CVs with typical-mass white dwarfs. In our case, the rareness of such systems may be attributed to the requirement of a small Roche-lobe offset $\Delta r_{\rm L}$, similarly to T Pyx and IM Nor (Section \ref{sec:offset}). 

\section{Summary}\label{sec:summary}

The donor stars in CVs are irradiated during novae by luminosities that exceed their own by several orders of magnitude. We revisited the response of the donor to this irradiation, focusing on the mass transfer $\dot{m}$ that is induced by the donor's inflation beyond its Roche lobe. We calculated the penetration of heat into the donor's atmosphere both analytically and using \textsc{mesa}. We coupled the heating to the mass transfer and derived $\dot{m}\propto m^{1.3}T_{\rm irr}^{5/3}$, where $m$ is the donor's mass, and $T_{\rm irr}$ is the irradiation temperature on its surface. 
Our calculation improves upon previous studies \citep{Kovetz88,Hillman2020} by consistently computing the mass and density of the heated zone that overflows the Roche lobe, rather than assuming a constant photospheric density. We find mass transfer rates that are more than an order of magnitude higher than previously thought --- $\dot{m}\sim 10^{-6}\msunyr$ at the peak of the nova eruption.

As the nova subsides and $T_{\rm irr}$ declines, $\dot{m}$ drops to lower values. However, we identified a self-sustaining state in which the white dwarf's accretion luminosity itself is sufficient to drive mass transfer by irradiation at a rate of $\dot{m}\sim 10^{-7}\msunyr$ long after the eruption. This `bootstrapping' state is manifested differently depending on the CV's orbital period (which is linked to the donor's mass through the Roche-lobe filling condition). For a typical CV white dwarf with $M\approx 0.8\,{\rm M}_\odot$ \citep{Zorotovic2011}, the critical period that determines the nature of the self-sustaining state coincides with the famous $\approx 2-3$ h period gap \citep{Knigge2011}:
\begin{enumerate}
    \item
    Short-period CVs erupt as recurrent novae during bootstrapping. Bootstrapping may explain the high accretion rates measured for the recurrent novae T Pyx and IM Nor between eruptions, as well as their short recurrence times \citep{Patterson2017,Patterson2020,Godon2018}.
    \item
    Long-period CVs burn hydrogen stably during bootstrapping and appear as supersoft X-ray sources then. Bootstrapping may explain the prolonged \citep[longer than the several years that are expected post nova, see][]{Soraisam2016} supersoft emission from RX J0537.7--7034 and 1E 0035.4--7230. With orbital periods of 3.5 and 4.1 h, these objects defy the standard explanation which invokes mass transfer on a thermal time-scale form especially massive donors \citep{King2001}.
\end{enumerate}

The proximity of the self-sustaining $\dot{m}$ to the hydrogen burning stability limit (Fig. \ref{fig:MWD}) motivates a more detailed theoretical or observational calibration of our spherically symmetric estimates, in order to determine exactly which systems are expected to appear as recurrent novae and which as supersoft sources.

Over the course of $\sim 10^3$ yr, stable mass transfer at this high rate increases the donor's Roche lobe, lowering $\dot{m}$ and $T_{\rm irr}$, until the star no longer overflows its Roche lobe --- bootstrapping ceases and $\dot{m}$ plummets. We compared this time-scale to the rate at which gravitational waves and magnetic braking tend to shrink the Roche lobe and estimated that only about $10^{-3}$ of CVs are currently in the bootstrapping state.
Nonetheless, novae recur $\sim 10^3$ times more often during bootstrapping \citep[thanks to the higher accretion rate, see][]{Chomiuk2020}, such that a significant fraction of nova eruptions may evolve to this state as the white dwarf cools --- broadly consistent with the occurrence of T Pyx and IM Nor. 

We conclude that novae trigger episodes of fast mass transfer by irradiation: $\dot{m}\sim 10^{-6}\msunyr$ at the peak of the eruption and, under certain circumstances (Section \ref{sec:offset}), a self-sustaining $\dot{m}\sim 10^{-7}\msunyr$ that persists long after the eruption subsides. The occurrence rate and duration of the self-sustaining phase depend critically on how the donor's Roche lobe evolves over time and through multiple nova cycles. If all donor stars and all novae were identical, all systems at a given orbital period and with a given white dwarf mass would evolve the same way after a nova: either into a self-sustaining state or not, depending on the Roche lobe's size during quiescence and during an eruption. This is inconsistent with the fact that systems at the same orbital period appear to have quite different accretion rates years after novae. We speculate that this difference is due to variation in either the surface layer structure of the donor or the mass and angular momentum lost during novae \citep[e.g.][]{Salazar2017,Schaefer2020,Chomiuk2020}, both of which are critical for determining if the self-sustaining state is realized. Because of these uncertainties, we cannot robustly test our predictions using the fraction of systems that appear to be in the self-sustaining state. Nevertheless, the bootstrapping mechanism naturally provides a unified explanation for two puzzling classes of close binaries: long-lived supersoft sources with short orbital periods, and recurrent novae with even shorter periods.    

\section*{Acknowledgements}

We thank Yael Hillman, Brian Metzger, Bradley Schaefer, and Ken Shen for comments and for illuminating discussions. We also thank the anonymous reviewer for a thoughtful report which has improved the paper. SG is supported by the Heising-Simons Foundation through a 51 Pegasi b Fellowship. This work benefited from workshops supported by the Gordon and Betty Moore Foundation through Grant GBMF5076.

\section*{Data Availability}

The data underlying this article will be shared on reasonable request to the corresponding author.



\bibliographystyle{mnras}
\input{nova.bbl}



\appendix

\section{Bootstrapping solution stability}\label{sec:stability}

Formally, equation \eqref{eq:Delta_Tirr} has two solutions for low $\Delta r_{\rm L}$, but only the stable one is plotted in Fig. \ref{fig:delta_rl}. If we define $\Tilde{T}\equiv T_{\rm irr}/T_{\rm irr}^0$ and $\Delta\equiv\Delta r_{\rm L}/h(T_{\rm irr}^0)$, then the equation reads $y=\Tilde{T}^{7/2}+\Delta/\Tilde{T}-1=0$. From the derivation of equation \eqref{eq:Delta_Tirr} it is understood that $y$ measures the difference between $\dot{m}$ that is required to sustain an irradiation temperature of $\tilde{T}$ by accretion and $\dot{m}$ that is driven out of the donor by such irradiation. A positive (negative) $y$ therefore decreases (increases) $\Tilde{T}$. From here it is easy to see that the high $\Tilde{T}$ solution (including our original solution for $\Delta=0$) is stable whereas the low one is unstable.

The same technique can be used to assess the stability of the bootstrapping solution to large fluctuations in $\dot{m}$ (and therefore $\Tilde{T})$, as observed for polars \citep[AM Herculis stars, e.g.][]{WuKiss2008}. We find that any increase in $\Tilde{T}$ is stable, but if $\Tilde{T}$ decreases below the low $\Tilde{T}$ solution, then $\Tilde{T}$ plummets further unstably and bootstrapping shuts off. Quantitatively, in the limit of $\Delta\ll 1$, the two solutions are $\Tilde{T}\approx \Delta \ll 1$ and $\Tilde{T}\approx 1$, such that the stable (high $\Tilde{T}$) solution retains its stability unless $\dot{m}$ drops by several orders of magnitude (remember that $\dot{m}\propto \Tilde{T}^4$), which is more than polars fluctuate. For higher values of $\Delta$, the two solutions get closer (until they coincide for $\Delta\approx 0.51$) which means that progressively smaller fluctuations are able to shut off bootstrapping.


\bsp	
\label{lastpage}
\end{document}